\documentclass[reprint,amsmath,amssymb, aps,]{revtex4-1}
\usepackage{graphicx}
\usepackage{dcolumn}
\usepackage{bm}
\usepackage{hyperref}

\begin{document}
\title{Angular Momentum Fluctuations of a Schwarzschild Black Hole}
\author{Marcelo Schiffer}
\affiliation{Department of Physics, Ariel University, Ariel 44837, Israel.}

\date{\today}

\begin{abstract}

In this paper we consider  angular momentum fluctuations of a Schwartzschild black hole in thermal equilibrium with radiation which, for the sake of simplicity is here modeled  by a scalar field.   Important, we do not set the black hole angular momentum $J$   identically  to zero at the outset; we allow it to have
a small value (in the sense that $J/M<<1$) and  then study  the  conditions for thermodynamical equilibrium; only then take the $J\rightarrow 0$ limit.  We calculate the black hole's angular momentum fluctuations  which turn out   to have two independent contributions: one that comes from the black hole itself, with no respect to the radiation, and another one which arises from the radiation. The result is astonishingly simple: the radiation contribution  depends exclusively on the cut-off proper distance from the horizon (or equivalently, the width of the brick-wall), while the black hole contribution is proportional to its event horizon  area. Accordingly, there are no strictly static black holes in nature, they randomly rotate in all possible  directions.  Since a black hole is nothing but geometry, we are dealing with geometry fluctuations -- our results are of quantum-gravitational nature (albeit at semi-classical level). Interestingly enough, if we apply  to the black hole  fluctuations component the (quantum) rules of angular momentum we obtain an event horizon area quantization rule, albeit  with a  different spectrum from equally spaced area spectrum which is widely accepted  in the literature.
 \begin{description}
\item[PACS numbers : 04.70Dy,  05.40.-a,  05.70-a,52.25 Kn]
\end{description}
\end{abstract}
\maketitle
\begin{widetext}
\section{A static black hole in equilibrium with radiation}

A rotating black hole of mass $M$ and angular momentum $J$ is described by the  Kerr geometry. In  Boyer-Lindquist coordinates the metric coefficients take the form:
\begin{eqnarray}
g_{tt} &=&-\left(1-\frac{ 2 M r}{\Sigma} \right)\, ,\\
 g_{\varphi\varphi}&=&\left(r^2+a^2 +2 \frac{M a^2 r \sin^2 \theta}{\Sigma}\right) \sin^2 \theta\, , \\
g_{t \varphi }&=&-\frac{2 M a r \sin^2\theta}{\Sigma}\, ,\\
g_{\theta\theta}&=&\Sigma\, ,  \\
g_{rr}&=&\frac{\Sigma}{\Delta}\, ,
\end{eqnarray}
where
\begin{eqnarray}
\Sigma&=&r^2+a^2 \cos^2 \theta\, ,\\
\Delta& =&r^2-2M r +a^2 \, ,
\end{eqnarray}
with $a=J/M$. The black hole event horizon is located at 
\begin{equation}
r_+=M+\sqrt{M^2-a^2}
\end{equation}
and the corresponding  area is
\begin{equation}
A=8\pi M r_+\, .
\end{equation}

In this geometry, particles are dragged  with angular velocity $\Omega=-g_{\varphi t}/g_{\varphi\varphi}$. Now consider such a black hole in thermal equilibrium with a bath of scalar particles within a confining vessel of radius $R>>2M$,  observed by an an observer rotating with a constant   angular velocity $\omega$ . The need of a rotating system will become clear shortly. The total energy and angular momentum of the system are
\begin{eqnarray}
{\cal E}&=&M+\sum_m\int \sqrt{-g} dr d\theta d\varphi  \int \frac{n(\epsilon,m,r,\theta) \epsilon  }{e^{\beta (\epsilon -m\hbar \omega)}-1}d\epsilon\label{energy}
\, ,\\
{\cal J}&=& J+\sum_m \int \sqrt{-g} dr d\theta d\varphi\int \frac{n(\epsilon,m,r,\theta) m\hbar  }{e^{\beta (\epsilon -m\hbar \omega)}-1} d\epsilon\label{angularm}\, ,
\end{eqnarray}
where $n(\epsilon,m,r,\theta)$ represents the density of states per unit volume with a fixed azimuthal quantum number $m$ in the rest frame.  The free energy of the radiation is:
\begin{equation}
\beta F=\sum_m \int  \sqrt{-g} dr d\theta d\varphi \int n(\epsilon,m,r,\theta) \ln\left(1-e^{-\beta( \epsilon-\hbar m  \omega) }\right) d\epsilon \, .
\label{defF}
\end{equation}

We assume that the total angular momentum ${\cal J}$ vanishes, the black hole's angular momentum fluctuations result from the absorption and emission of quanta from the radiation. The  angular momentum conservation condition (eq.(\ref{angularm})) can be expressed in terms of the free energy as:
\begin{equation}
J -\left(\frac{\partial F}{\partial \omega}\right)=0 \, .
\label{J}
\end{equation}

Similary,  the energy conservation  condition (eq. (\ref{energy}) )reads \cite{landaumechanics}:
\begin{eqnarray}
{\cal E}=M+ \omega J +\left(\frac{\partial(\beta F)}{\partial \beta}\right )\, .
\label{E}
\end{eqnarray}

The total entropy is given by
\begin{equation}
{\cal S}=\frac{2\pi M r_+}{\hbar}+\beta^2 \left(\frac{\partial F}{\partial \beta }\right)\, .
\end{equation}

The first term in the last equation corresponds to the  Bekenstein-Hawking entropy. Integrating  eq.(\ref{defF}) by parts
\begin{equation}
 F=-\sum_m \int \int \sqrt{-g} dr d\theta d\varphi \frac{\gamma(\epsilon,m,r,\theta)}{ e^{\beta (\epsilon-m\hbar  \omega)}-1} d\epsilon \, ,
\label{F}
\end{equation}
where $\gamma(\epsilon,m,r,\theta)$ is the total number of states (per  unit volume) for a given energy and azimuthal number $m$,  $n=d\gamma/d\epsilon$.
In the rotating frame, the azimuthal angle is $\tilde{\varphi}=\varphi -\omega t$ and the metric elements reads
\begin{eqnarray}
\tilde{g}_{tt} &=& g_{tt}+\omega^2 g_{\varphi\varphi} +2\omega g_{t\varphi} \label{gtt} \, ,\\
\tilde{g}_{t\varphi}&=& g_{t\varphi}+\omega g_{\varphi\varphi}\, ,
 \end{eqnarray}
all other metric coefficients remaining the same as the in the non-rotating frame. The energy measured in this frame is $\tilde{\epsilon}=\epsilon -m\hbar \omega$.  

In what follows, for the sake of completeness we follow the discussion given by  Chang-Young et all  \cite{Chang}. A massless scalar field  satisfies the wave equation :
\begin{equation}
\frac{1}{\sqrt{-\tilde{g}} }\partial_a \left(\tilde{g}^{ab} \sqrt{-\tilde{g}} \partial_b \Phi\right) -\xi R\Phi =0\, ,
\end{equation}
where $\xi$ represents the coupling constant. Neglecting back-reaction of the geometry and a  semi-classical approximation
$\Phi =\Phi_0 e^{i(-\tilde{\epsilon} t + m\tilde{\varphi} +S(r,\theta))/\hbar}$. Then, it follows that 
\begin{equation}
\tilde{\epsilon}^2 \frac{\tilde{g}_{\tilde{\varphi}\tilde{ \varphi}}}{D}+2\frac{\tilde{g}_{\tilde{\varphi} t} \tilde{\epsilon}}{D} m -m^2 \frac{(-\tilde{g}_{tt})}{D}-\left[ \frac{1}{g_{rr}}  p_r^2 +\frac{1}{g_{\theta\theta }}  p_\theta^2\right]=0 \, ,
\label{wkb}
\end{equation}
with $p_r=\frac{\partial S}{\partial r}$, $p_\theta=\frac{\partial S}{\partial \theta}$,  $D=-g_{rr}\tilde{g}_{tt}+\tilde{g}_{t\tilde{\varphi}}^2$ and we inverted also the metric In the semi-classical approximation the number of states  for a fixed value of $m$ in the rotating frame is the volume in phase space \cite{thooft}-\cite{minho} :
\begin{equation}
\gamma(\tilde{\epsilon},m)=\frac{1}{(2\pi \hbar)^3}  \int d\tilde{\varphi} \,   dp_\theta d\theta \, dp_r \,dr \, .
\label{gamma} 
\end{equation}
Performing the  immediate integrations over $p_r$ and $\tilde{\varphi}$, and  inserting the value of $p_r(r,\theta)$ obtained from eq. (\ref{wkb})
\begin{equation}
\gamma(\tilde{\epsilon},m)=\frac{1}{(2\pi^2 \hbar)^3} \int d\tilde{\varphi} dp_\theta d\theta \, \,dr \sqrt{g_{rr} }\left[ \frac{ \tilde{g}_{\tilde{\varphi} \tilde{\varphi}}\tilde{\epsilon}^2+2\tilde{g}_{t\tilde{\varphi}} m \tilde{\epsilon}-m^2(-\tilde{g}_{tt})}{D}-
\frac{p_\theta^2}{g_{\theta\theta}}\right]^{1/2}\, .
\end{equation}
Then, integrating over the classically allowed region it follows that
\begin{equation}
 \gamma(\tilde{\epsilon},m)=\frac{1}{16\pi^2 \hbar^3} \int  d\theta \,dr \frac{\sqrt{g_{rr} g_{\theta\theta}}}{D}\left(  \tilde{g}_{\tilde{\varphi} \tilde{\varphi}}\tilde{\epsilon}^2+2\tilde{g}_{t\tilde{\varphi}} m \tilde{\epsilon}-m^2(-\tilde{g}_{tt})
\right)\, .
 \end{equation}
  
The total number of states for all $m$ is  given by summing  $\sum_m \gamma(\tilde{\epsilon},m)$ . Approximating the sum over $m$ by an   by integral over the region where the integrand is positive yields
\begin{equation}
\gamma(\tilde{\epsilon})=\frac{1}{12\pi^2 \hbar^3} \int \sqrt{-g } d\tilde{\varphi}d\theta \,dr \frac{1}{(- \tilde{g}_{tt})^2} \tilde{\epsilon}^3\, ,
\label{gamma}
 \end{equation}
 where $-g=g_{rr} g_{\theta\theta}D$ is the determinant of the Kerr metric.

 Inspecting this expression, it is easy to identify the density of states in the rotating frame.
 \begin{equation}
\gamma(\tilde{\epsilon},r,\theta)=\frac{1}{12\pi^2 \hbar^3(- \tilde{g}_{tt})^2}  \tilde{\epsilon}^3 \, ,
 \end{equation}
Inserting this density of states in eq.(\ref{F}) and performing  the Bose-Einstein-like integration , the free energy boils down to a simple result
\begin{equation}
 F=-\frac{\pi^2 }{120 \hbar^3 \beta^4} V \, ,
\label{Ffinal}
 \end{equation}
 where we defined
 \begin{equation}
V=  \int_{r_++h}^R dr \int  d\theta\frac{\sqrt{-g}}{\tilde{g}_{tt}^2}\, ,
\label{V}
\end{equation}
with
 \begin{equation}
 \tilde{g}_{tt}=g_{tt}+(\omega^2  -2\omega \Omega) g_{\varphi\varphi} \, .
 \end{equation}
 Inspecting eqs. (\ref{Ffinal}),(\ref{V}) we notice that the relevant (inverse) temperature in the free function is is  $\beta \sqrt{-g_{tt}}$   which is nothing but Tolman's inverse temperature, the local temperature measured by the rotating observer \cite{tolman}-\cite{mattvisser}. 
Following t'Hooft, we introduced a cut-off $h$ at the horizon that can either represents a brick wall $\Phi ( r_+ +h)=0 $ or  our ignorance of how to properly renormalize the divergences at the horizon.  The total energy  reads 
\begin{equation}
{\cal E}=M+\omega J+\frac{\pi^2 }{40 \hbar^3 \beta^4} V
\end{equation}
where $\omega$  is to be regarded as a  chemical potential that implements angular momentum conservation  (eq.(\ref{J})):
\begin{equation}
\frac{30 \hbar^3 \beta^4}{\pi^2}J+  \int_{r_++h}^R dr \int  d\theta \frac{\sqrt{-g}}{(-\tilde{g}_{tt})^3} g_{t\varphi}
+\omega \int_{r_++h}^R dr \int  d\theta \frac{\sqrt{-g}}{(-\tilde{g}_{tt})^3 }g_{\varphi\varphi}=0 \, .
\label{findw}
\end{equation}
Equivalently $\omega$ can be regarded as the angular velocity of a rotating observer must have such that the  total angular momentum vanishes in his frame.
 
 At last, the total entropy reads
\begin{equation}
{\cal S}= \frac{4\pi M r_+}{\hbar}+\frac{\pi^2}{30\hbar^3 \beta^3} V \, .
\label{S}
\end{equation}
The first derivatives $(\frac{\partial {\cal S}}{\partial M})_J=0$ together with the energy constraint (eq.(\ref{E})) gives the radiation temperature 
\begin{equation}
\beta= \frac{4\pi M }{\hbar}\left( \frac{M}{\sqrt{M^2-J^2/M^2}}+1\right)\, .
\end{equation}
In order to study thermodynamical fluctuations of a Schwatzschild  we need to expand the entropy up to the second order in $J$ (or $a$). At the lowest order in  $J$:
\begin{equation}
g_{t\varphi}\approx -\frac{2J \sin^2 \theta}{r} \, ,
\end{equation}
so the second term in eq.(\ref{findw}) is at least linear in $J$, and  so must be also $\omega$.
Solving this equation at the first order in $J$,  we replace the Kerr metric coefficients by  the Schwartzschild metric with the exception of  $g_{t\varphi}$,  which takes the above value. Then :
\begin{equation}
\frac{30 \hbar^3 \beta_0^4}{\pi^2} J-2 J \int d\theta \sin^3\theta \int_{2M+\delta}^R \frac{r^4 dr}{(r-2M)^3}
+\omega \int d\theta \sin^3\theta \int_{2M+\delta}^R \frac{r^7 dr}{(r-2M)^3}=0 \, .
\end{equation}
were $\beta_0=8\pi M/\hbar$ . The density of states is highly peaked near the horizon \cite{Chang}, so most of the contribution to the integral comes from the lower integration limit. After some algebra
\begin{equation}
J=4M^3  \left[1+ 3\left(\frac{\delta}{M}\right) +{\cal O} \left(\frac{\delta}{M}\right)^2 \log \left(\frac{\delta}{M}\right)\right]\omega\, .
\end{equation}
 Notice that 
\begin{eqnarray}
r_+&\approx& 2M -\frac{a^2}{2M} \label{rplus}\\
\beta^{-3}&\approx& \left(\frac{\hbar}{8\pi M}\right)^2 \left(1-\frac{3a^2}{4M^2}\right) \, .
\label{approxb}
\end{eqnarray}
Expanding eq.(\ref{S}) to the second order in $a$ is a bit more sweaty. Let $f(r,\theta)$ represent the integrand in  eq.(\ref{V}),then
\begin{equation}
\int_{r_+ + \delta}^R f(r,\theta) drd\theta =  \int_{2M+\delta}^R f_0(r,\theta)dr d\theta +
 \int_{2M+\delta-a^2/2M}^{2M+ \delta} f_0(r,\theta) dr d\theta 
 + \int_{2M+\delta}^R f_2(r,\theta)dr d\theta \, ,
\end{equation} 
where $f_0(r,\theta),f_2(r,\theta)$ represent the zeroth and second order expansion in terms of $a^2$.  Since 
\begin{equation}
\sqrt{-g}=(r^2+a^2 \cos^2 \theta)\sin\theta \, ,
\end{equation}
and at the relevant order
\begin{equation}
\Omega \approx \frac{2Ma}{r^3}\, ,
\end{equation}
then
\begin{equation}
\frac{1}{(\tilde{g}_{tt})^2}\approx \frac{r^2}{(r-2M)^2} - \frac{4 M a^2 \cos^2\theta}{(r-2M)^3} + 2\omega^2\frac{r^5\sin^2\theta}{(r-2M)^3}
-\omega\frac{8 M a r^2\sin^2\theta}{(r-2M)^3} \, .
\end{equation}
At last,
\begin{equation} 
V=V_0 -\frac{a^2}{M} \frac{(2M+\delta)^4}{\delta^2}+\frac{2a^2}{3} \int_{2M+\delta}^R \frac{r^2 dr}{(r-2M)^2}-\frac{8 Ma^2}{3} \int_{2M+\delta}^R \frac{r^2 dr}{(r-2M)^3}+\frac{8\omega^2}{3}\int_{2M+\delta}^R \frac{r^7 dr}{(r-2M)^3}
-\frac{32M a \omega}{3} \int_{2M+\delta}^R \frac{r^4 dr}{(r-2M)^3}
\label{approxV}
\end{equation} 
where 
\begin{equation}
V_0=2\int_{2M+\delta}^R \frac{r^4 \sin \theta dr d\theta}{(r-2M )^2}
\end{equation}
The density of states is very very large near the horizon \cite{Chang,minho}   and we take only the contribution from the lower limit of the integral where the divergence occurs.  Putting all the pieces  together (eqs.(\ref{S},\ref{rplus},\ref{approxb}\ref{approxV}) the total entropy boils down to
\begin{equation}
{\cal S}={\cal S}_0 - \frac{2\pi J^2}{\hbar M^2}-\frac{J^2}{480\pi M^2 \delta^2}-\frac{J^2}{480\pi M^3 \delta} \, ,
\end{equation}
where ${\cal S}_0$ is the total entropy for $J=0$. Clearly the equilibrium condition $(\frac{\partial S}{\partial J})_M=0$ is satisfied identically. Angular momentum fluctuations are given by \cite{landaulifshitz} :
\begin{equation}
\frac{1}{(\Delta J)^2}=-\left(\frac{\partial^2 {\cal S}}{\partial J^2}\right)_{J=0}
\end{equation}
Keeping only the most divergent term and expressing the  coordinate distance in terms of the proper distance $\delta=\Delta^2/8M$, we can write
\begin{equation}
\frac{1}{(\Delta J)^2}=\frac{1}{(\Delta J)_{BH}^2}+\frac{1}{(\Delta J)_{\mbox{field}}^2}\, ,
\end{equation}
where these terms represent black hole and radiation fluctuations:
\begin{equation}
(\Delta J)_{BH}= \sqrt{\frac{\hbar }{4\pi }}M\quad \quad , \quad \quad (\Delta J)_{field}=\frac{\sqrt{15 \pi }}{2}\Delta^2
\end{equation}

This is an amazingly simple result. Part of the fluctuations has origin in the black hole's quantum atmosphere, which is the (quantum) field within a small shell of proper-width $\Delta$ around the horizon. The black hole fluctuations are a bit  mysterious. Mathematically it originates from the Bekenstein-Hawking entropy - thus being a true property of the black hole. Let us consider only this angular momentum fluctuation. Assume $\overline{m}=0$ , then, from the basic properties of angular momentum  
\begin{equation}
\Delta m^2=\overline{m^2} = \frac{1 }{2l+1} \sum_{m=-l}^l m^2 \approx \frac{ l^2}{3}\hbar^2\, ,
\end{equation}
were we approximated the sum by an integral. Equating $\Delta J_{BH}^2=\hbar^2 \Delta m^2$ , if follows that the event horizon is quantized:
\begin{equation}
A_{BH}\approx  \frac{64 \pi^2 \hbar}{3}l^2\, . 
\end{equation}
This result is at odds with the linear spacing area spectrum vindicated by most authors \cite{bhspectroscopy}-\cite{dreyer}. Anyway it is very surprising that the black hole area quantization results from the quantum rules of  angular momentum. 
\section{Concluding Remarks}
Angular momentum fluctuations emerge from a thin shell of Planckian width and from the black hole itself. .The former is Universal, does not depend on the black hole mass but only upon the width of this quantum atmosphere; the latter depends on the event horizon area. Being so different, they must have different physical origins. That is to say, a Schwatzschild black hole rotates randomly in all possible directions. The physical meaning of the fluctuations remains elusive. Surprisingly, the rules of angular momentum implies that  the event horizon is quantized and  a for large  mass, the mass spectrum depends (nearly) linearly  on the quantum number $l$ which also relates to the angular momentum fluctuations. Furthermore assuming that the  cut-off parameter is Planckian, we can write $\Delta^4=\mu \hbar^2/15\pi^2$  for $\mu$ a numerical value of order one. Accordingly  the (averaged) Cauchy horizon $\overline{r_-}=\overline{J^2}/2M^3 \sim \mu \hbar^2 /(8\pi(M^2+\mu \hbar)M) $ never  vanishes. Since a black hole is nothing but  geometry, our results should be regarded as being of (semi-classical ) quantum gravity nature.
\end{widetext}

 \begin{quotation}
\flushleft
\bibitem{landaumechanics} L. D. Landau and E. M. Lifshitz (1970), {\it  Mechanics}, Pergamon Press
\bibitem{thooft} G. ’t Hooft, "On the quantum structure of a black hole' "Nucl. Phys. B 256, 727 (1985).
\bibitem{Chang}, "Rotating black hole entropy from two different viewpoints", Chang-Young Ee, Daeho Lee, M. Yoon, Classical and Quantum Gravity (2008).
\bibitem{minho} Min-Ho Lee* and Jae Kwan Kim ,"Entropy of a quantum field in rotating black holes", Phys. Rev. D 54, 3904 (1996)
\bibitem{tolman} R. C. Tolman (1930), “On the weight of heat and thermal equilibrium in general relativity,”
Phys. Rev. 35, 904 .
 \bibitem{MTW} C.W. Misner ; K.S. Thorne and J.A. Wheeler in {\it Gravitation}: Freeman 1973.
 \bibitem{mattvisser} Jessica Santiago and Matt Visser , {\it "Gravity’s universality: The physics underlying Tolman temperature gradients"}, International Journal of Modern Physics D, Vol. 27, No. 14, 1846001 (2018).
 \bibitem{hawking} S. W. Hawing, "Black Holes and Thermodynamics" (1976), Phys. Rev D13,191.
 \bibitem{parentani}{R Parentani, J Katz, I Okamoto,  "Thermodynamics of a black hole in a cavity Classical and Quantum Gravity" - iopscience.iop.org (1995).}
 \bibitem{takagi} Shin·ichi Takadi and Shin Takagi (1985), "Phenomenological Theory of the Equilibrium State of the System with a Black Hole and Radiation",   Prog. Theor. Phys. Vol. 73, No.1,288.
 \bibitem{huang} Kerson Huang, {\it Statistical Mechanics} (1987), John Willey.
 \bibitem{pavon1}{General Relativity and Gravitation ,D. Pavon, W. Israel , Vol. 16, Issue 6, pp 563-568 (1984).}
 \bibitem{pavon2}{D Pavon, J.M. Rubi,  " On some properties of the entropy of a system containing a black hole", General Relativity and Gravitation,  - Springer (1985).}
  \bibitem{landaulifshitz} L. D. Landau and E. M. Lifshitz (1970), {\it Statistical Physics}, Pergamon Press.
  \bibitem{bhspectroscopy} J. D. Bekenstein and V. F. Mukhanov,  "Spectroscopy of the quantum black hole", Phys. Lett. B360 (1995) 7 [gr-qc/9505012] 
  \bibitem{hod} S. Hod, "Bohr’s correspondence principle and the area spectrum of quantum black holes", Phys.
Rev. Lett. 81 (1998) 4293 .
  \bibitem{davidson}A. Davidson, "Holographic Shell Model: Stack Data Structure inside Black Holes?", Int. J. Mod.
Phys. D23 (2014) 1450041 [1108.2650].
\bibitem{dreyer} O. Dreyer, "Quasinormal modes, the area spectrum, and black hole entropy", Phys. Rev. Lett. 90
(2003) 081301 [gr-qc/0211076].
\end{quotation}

\noindent
\end{document}